\documentclass[10pt]{article}

\newcommand{\bra}[1]{\langle #1|}
\newcommand{\ket}[1]{|#1\rangle}

\usepackage[margin=2.8cm, a4paper]{geometry}
\usepackage{amsmath,amsthm,amsfonts,amssymb}
\usepackage{fontenc}
\usepackage[all]{xy}
\usepackage{graphicx,subfigure,multirow}
\usepackage[figurename=Fig.]{caption}
\usepackage{tikz}
\usetikzlibrary{shapes, arrows, shadows}
\usepackage[latin1]{inputenc}
\usepackage[scaled]{helvet}
\usepackage[toc, page]{appendix}
\usepackage{dsfont}

\numberwithin{equation}{subsection}
\usepackage{cancel}

\usepackage[colorinlistoftodos]{todonotes}

\begin{document}

\begingroup
\centering
{\Large\textbf{Higher-order interference in extensions of quantum theory} \\[1.5em]
  \normalsize Ciar\'{a}n M. Lee\textsuperscript{\dag,}\footnote{
Electronic address: ciaran.lee@cs.ox.ac.uk}  and John H. Selby\textsuperscript{\dag,*,}\footnote{Electronic address: john.selby08@imperial.ac.uk}}\\[1em]
\centering
\it \textsuperscript{\dag}  University of Oxford, Department of Computer Science, Oxford OX1 3QD, UK. \\ \it \textsuperscript{*} Imperial College London,  London SW7 2AZ, UK.

\endgroup



\begin{abstract}
Quantum interference, manifest in the two slit experiment, lies at the heart of several quantum computational speed-ups and provides a striking example of a quantum phenomenon with no classical counterpart. An intriguing feature of quantum interference arises in a variant of the standard two slit experiment, in which there are three, rather than two, slits. The interference pattern in this set-up can be written in terms of the two and one slit patterns obtained by blocking one, or more, of the slits. This is in stark contrast with the standard two slit experiment, where the interference pattern \emph{cannot} be written as a sum of the one slit patterns. This was first noted by Rafael Sorkin, who raised the question of \emph{why} quantum theory only exhibits irreducible interference in the two slit experiment. One approach to this problem is to compare the predictions of quantum theory to those of operationally-defined `foil' theories, in the hope of determining whether theories that do exhibit higher-order interference suffer from pathological -- or at least undesirable -- features. In this paper two proposed extensions of quantum theory are considered: the theory of Density Cubes proposed by Daki{\'c}, Paterek and Brukner, which has been shown to exhibit irreducible interference in the three slit set-up, and the Quartic Quantum Theory of {\.Z}yczkowski. The theory of Density Cubes will be shown to provide an advantage over quantum theory in a certain computational task and to posses a well-defined mechanism which leads to the emergence of quantum theory -- analogous to the emergence of classical physics from quantum theory via decoherence. Despite this, the axioms used to define Density Cubes will be shown to be insufficient to uniquely characterise the theory. 
In comparison, Quartic Quantum Theory is a well-defined theory and we demonstrate that it exhibits irreducible interference to all orders. This feature of {\.Z}yczkowski's theory is argued not to be a genuine phenomenon, but to arise from an ambiguity in the current definition of higher-order interference in operationally-defined theories. Thus, to begin to understand why quantum theory is limited to a certain kind of interference, a new definition of higher-order interference is needed that is applicable to, and makes good operational sense in, arbitrary operationally-defined theories.
\end{abstract}

\section{Introduction}

\subsection{Overview of results}


The present paper investigates two proposed extensions of quantum theory from the point of view of their interference behaviour. This investigation clarifies the impact of these two generalised theories to ongoing experimental tests for ``higher-order interference'' and explores potential information-theoretic consequences of post-quantum interference in concrete theories. In particular it highlights an ambiguity in the current definition of higher-order interference. The two theories which shall be investigated are: the theory of Density Cubes proposed by Daki{\'c}, Paterek and Brukner, which has been shown to exhibit third-order interference in the three slit set-up \cite{DensityCube}, and the Quartic Quantum Theory \cite{zyczkowski2008quartic} of {\.Z}yczkowski.

The five main conclusions of our investigation are as follows:  
\begin{enumerate}
\item The theory of Density Cubes posses a well-defined mechanism which leads to the emergence of quantum theory -- analogous to the emergence of classical physics from quantum theory via decoherence.
\item The theory of Density Cubes provides an advantage over quantum theory in a computational task based on the collision problem.
\item The axioms used to define the theory Density Cubes are insufficient to uniquely characterise it. It should hence be thought more as a framework for possible theories than a unique theory.
\item Quartic Quantum Theory (QQT) exhibits irreducible interference to all orders relative to the definition of higher-order interference provided by Barnum et al. in \cite{barnum2014higher}. \label{4}
\item Point \ref{4}, above, explicitly highlights an ambiguity in the current definition of higher-order interference which must be taken into account in future experimental investigations of higher-order interference.
\end{enumerate}

The rest of this paper is organised as follows. In the next section we motivate the study of higher-order interference, present some of the previous literature on this topic, and discuss our main results in more detail. In section \ref{GPT} we provide a definition of higher-order interference and discuss an operational framework of hypothetical physical theories in which such a definition can be rigorously explored. In section \ref{DC}, the results concerning the theory of Density Cubes shall be presented. In section \ref{QQT}, the result concerning Quartic Quantum Theory shall be shown.

\subsection{Background and motivation}

The predictions of quantum theory are the most accurately tested of any physical theory in the history of science. Nevertheless, it may turn out to be the case that quantum theory is only an effective description of a more fundamental theory whose predictions deviate from those of quantum theory in certain energy regimes or sufficiently sensitive experimental set-ups. It is thus of the utmost importance that fundamental tests of the validity of quantum theory be performed. Such tests take a characteristically quantum prediction and probe the limits of its accuracy in different experimental situations. One such prediction, currently under experimental investigation \cite{sinha2008testing,park2012three}, is the limitation of quantum theory to second, as opposed to higher, order interference in $n$-slit experiments.

Higher-order interference was first described by Sorkin \cite{sorkin1994quantum} who noted that quantum theory is limited to having only second-order interference. Informally, this means that the interference pattern obtained in a three -- or more -- slit experiment can be written in terms of the two and one slit interference patterns that are obtained by blocking some of the slits. Thus there are no genuinely new features resulting from considering three slits instead of two. This is in stark contrast with the existence of second-order, i.e. quantum-like, interference, for which there exists a two-slit experiment whose interference pattern \emph{cannot} be written as a sum of the one slit patterns obtained by blocking each one of the slits. This was first made precise in the context of quantum measure theory \cite{sorkin1995quantum}, where moving from classical to quantum theory can be seen as a weakening of the Kolmogorov sum rule to allow for second (but not third, or higher) order effects.
 
Restriction to only second-order interference appears to be a characteristically quantum phenomena and many other `quantum-like' features can be derived from it. For example: limiting correlations \cite{dowker2014histories, niestegge2013three} to the `almost quantum correlations' discussed in \cite{MattyJ}, and bounding contextuality \cite{henson2014bounding}. 
Additionally a lack of third-order interference was also used by Barnum, M{\"u}ller and Ududec \cite{barnum2014higher} as a postulate in their reconstruction of quantum theory.

The natural question that arises from this discussion is \emph{why} does quantum theory only exhibit second-order interference? It may strike some as odd that there is a limit to the non-classicality of quantum theory. Why is nature strange, but not excessively so? Does the existence of genuine third-order interference violate some physical principle, such as non-signalling \cite{barrett2007information}, that we take to be fundamental? We do not fully answer this question here but, by investigating two operationally-defined hypothetical extensions of quantum theory, we gain some insight into possible consequences of the existence of post-quantum interference.

One way to approach this problem is to consider quantum theory in the context of a widely studied framework used to discuss possible extensions to the quantum formalism. This framework, known as the generalised probabilistic theory (GPT) framework \cite{barrett2007information, hardy2001quantum}, is general enough to accommodate essentially arbitrary operational theories, where an operational theory specifies a set of laboratory devices that can be connected together in different ways and assigns probabilities to different experimental outcomes. Theories in this framework have the minimal amount of structure required to provide a consistent account of arbitrary operational scenarios \cite{barrett2007information}. It should be noted however that it is not the purpose of theories in this framework to tell us how post-quantum physics could potentially arise, but to provide a consistent operational model who's predictions deviate from those of quantum theory. The only considerations of interest are those which are operational\footnote{Note that operationalism as a philosophical viewpoint, in which one asserts that there is no reality beyond laboratory device settings and outcomes, is not being espoused here. One should merely view the approach taken here as an operational methodology aimed at gaining insight into certain structural properties of physical theories.} in nature.


Barnum, M{\"u}ller and Ududec have provided an operational definition \cite{barnum2014higher} (see also \cite{ududec2011three}) of higher-order interference that is applicable to any GPT, which we review in scetion \ref{GPT}. Given this definition, one can attempt to construct a GPT that exhibits higher-order interference in the hope of using it as a `foil' to quantum theory. Such a foil theory would hopefully shed some light on possible pathological -- or at least undesirable -- features of higher-order interference theories and thus provide reasons `why', in some sense, quantum theory should be limited to second-order interference. Currently, to the best of the authors knowledge, there are no `complete' GPTs that exhibit third-order interference. There are particular state spaces \cite{UdudecThesis} that have higher order interference but these are of a fixed dimensionality and composition is not discussed, additionally they have a highly restricted set of dynamics when compared to quantum theory. 

It would be of particular interest if there was a theory that exhibited higher-order interference and which contained quantum theory as a limiting case. Yet, if such a theory exists, there should be some mechanism by which the magnitude of effects unique to this theory are suppressed, thus explaining why quantum theory is such a good effective description of the world. This mechanism would be analogous to the process of decoherence, which induces the quantum-classical transition and which makes observation of genuine quantum effects hard to experimentally detect. Therefore the mechanism by which an extension of quantum theory reduces to standard quantum theory is called \emph{hyper-decoherence}\footnote{See \cite{zyczkowski2008quartic} for a more in-depth discussion of hyper-decoherence.}. Any well-defined theory that extends quantum theory should provide a mechanism for hyper-decoherence. Experimental bounds have been found limiting the possible amount of third (or higher) order interference \cite{sinha2008testing,park2012three}, thus placing stringent bounds on the hyper-decoherence time of potential extensions of quantum theory.

Ududec, Barnum and Emerson have shown \cite{ududec2011three} that the absence of third-order interference is equivalent to the ability to perform full tomography of any state using only measurements consisting of two-slit experiments, i.e by only performing measurements on two dimensional subsystems\footnote{i.e. by only performing measurements of the form $a\bra{i}+b\bra{j}$.}. It follows that any theory which exhibits genuine third-order interference, and aims to be an extension of quantum theory, requires more parameters to specify an $n$-level system than are required to specify an $n$-level quantum system. Intuitively then, one can think of the dimension of the subspace upon which one needs to perform measurements to do complete tomography as corresponding to the order of interference.

Guided by this, Daki{\'c}, Paterek and Brukner \cite{DensityCube} have proposed a method to construct a theory that exhibits third-order interference and which extends standard quantum theory. In section~\ref{hyper}, we demonstrate that this construction gives rise to a sensible notion of hyper-decoherence which leads to the emergence of quantum theory in particular cases -- analogous to the emergence of classical physics from quantum theory. In section~\ref{Advantage} we also show that this construction provides an advantage over quantum theory in a certain computational task. Despite this, in section \ref{DC1} of this paper it will be shown that this approach -- as it is currently presented -- does not lead to a well-defined physical theory. We show that the axioms defining the state space are insufficient to uniquely characterise the theory. It is therefore suggested that one can view the theory of Density Cubes more as a framework for developing operational theories than a unique theory. Moreover, although non-trivial (non-quantum) transformations have been identified, these axioms allow for unphysical transformations that map physical states to states that give complex-valued probabilities on measurement.

Another feature of tomography in the GPT framework is discussed by Hardy in \cite{hardy2001quantum}, where a hierarchy of theories are presented and shown to satisfy the relation $K=N^r$, where $K$ is the number of effects whose statistics are required to completely determine a state, $N$ is the dimension of the system and $r$ is a positive integer specifying the level in the hierarchy. The case $r=1$ corresponds to classical theory and $r=2$ to quantum theory.\footnote{Note that we are allowing sub-normalised states, hence quantum theory satisfies $K=N^2$ rather than $K=N^2-1$.} For $r>2$ one may expect -- based on the results of \cite{ududec2011three} discussed above -- that tomography on these higher dimensional subspaces leads to higher-order interference. The results of \cite{ududec2011three} suggest that the $r$th level of this hierarchy, i.e. $K=N^r$, should exhibit $r$th-order interference, but no higher.

{\.Z}yczkowski has developed a theory \cite{zyczkowski2008quartic} satisfying $K=N^4$, which extends quantum theory, and so provides a candidate for a theory of higher-order interference. In section \ref{QQT} of this paper, it is shown that {\.Z}yczkowski's $K=N^4$ theory does not suffer from many of the problems of Daki{\'c} \emph{et al.'s} construction; there is a unique state space associated with the theory and all transformations are physical. Furthermore, this theory does indeed exhibit third -- and higher -- order interference. In fact, this theory exhibits $n$th-order interference for all $n$, which is somewhat surprising and unexpected, as one would expect, based on the discussion in the previous paragraph, that this theory exhibits at most $4$th-order interference. Another surprising, and somewhat worrying, feature of interference in this theory, that will be shown in section~\ref{interference}, is the fact that the existence of higher-order interference stems from a somewhat artificial and operationally unmotivated choice. Blocking some subset of the slits, which correspond to apertures in the physical barrier describing an $n$-slit experiment, should uniquely define a measurement, but {\.Z}yczkowski's theory does not posses this feature; there exist (at least) two well-defined measurements that correspond to blocking the same subset of slits in an $n$-slit experiment. One of these measurements results in higher-order interference, the other does not.

Arguably, both of these features arise from a limitation of Barnum, M{\"u}ller and Ududec's definition of higher order interference rather than a genuine phenomenon; there should be a unique measurement that corresponds to opening any subset of slits, and this does not appear to happen without further constraints on the theory.\footnote{It should be noted that all theories of interest to Barnum, M{\"u}ller and Ududec do satisfy these extra constraints, and so their definition suffices for all considerations of interest in \cite{barnum2014higher}.} Thus one should not consider {\.Z}yczkowski's theory as an example of higher-order interference in the sense originally meant by Sorkin, but rather a demonstration of the challenges of applying his original definition to arbitrary GPTs. Thus to begin to understand the reason why, in some sense, quantum theory is limited to second-order interference, we first need a definition of higher-order interference that is applicable to, and makes good operational sense in, arbitrary GPTs. Ways in which such a definition might arise are discussed in section \ref{con}.

Finally, in section~\ref{Correlation}, we briefly comment on the type and strength of correlations allowed in {\.Z}yczkowski's theory and provide evidence of a speed-up over quantum theory in communication complexity problems.

\section{A definition of higher-order interference in generalised probabilistic theories}   \label{GPT}

Any physical theory must provide a consistent explanation of experimental results. This provides the basic idea underlying the framework of generalised probabilistic theories, where the fundamental notions are operational in nature. Theories in this framework have the minimal amount of structure required to provide a consistent account of arbitrary operational and experimental scenarios. As discussed in the introduction, it should be noted that it is not the purpose of theories in this framework to tell us how post-quantum physics could potentially arise (via some high-energy probing of a certain field theory perhaps), but to provide a consistent operational model who's predictions deviate from those of quantum theory. The only considerations of interest are those which are operational in character.

A generalised probabilistic theory (GPT) specifies a set of laboratory devices that can be connected together in different ways and prescribes probabilities to different experimental outcomes. States, which correspond to experimentally preparing a system, and effects, which correspond to the outcome of some measurement on a system, are taken as primitive notions in the GPT framework. The set of states is know as the state space and the set of effects is known as the effect space. Transformations between different states are allowed, but it is demanded that all \emph{physically allowed} transformations map the state space to itself. For a review of the GPT framework, see \cite{barrett2007information, hardy2001quantum}.

Barnum, M{\"u}ller and Ududec \cite{barnum2014higher} have provided a definition of higher-order interference that is applicable to any GPT and is equivalent to Sorkin's original definition in the quantum and classical cases. This definition takes its motivation from the set-up of certain experimental interference experiments, in which a particle (a photon or electron, say) passes through apertures, which correspond to the slits, in a physical barrier. By blocking some of the slits and repeating the experiment many times, one can build up an interference pattern on a screen placed behind the physical barrier. For a more in depth discussion, see Section V. of \cite{barnum2014higher}.

The Barnum \emph{et al.} definition of higher-order interference proceeds as follows. They firstly define exposed faces, $F_i$, of the state space as a set of states for which there exists an effect\footnote{We are using curved rather than angular `Dirac' notation to denote states and effects in a GPT.} $(f_i|$ satisfying $(f_i|s)=1  \iff |s)\in F_i$. We should think of the effect $(f_i|$ as the effect corresponding to placing a detector just behind the slit $i,$
the face $F_i$ is therefore the set of states that are detected at slit $i$ with certainty. The union of exposed faces is defined, $F_{ij}:=F_i\cup F_j$ as the smallest exposed face that includes both $F_i$ and $F_j$, this is the face generated by an effect that is a coarse graining of the effects behind $i$ and $j$. Faces are disjoint $F_i\perp F_j$ if $(e_i|s)=0, \forall |s)\in F_j$ and $(e_j|s)=0, \forall |s)\in F_i$. We expect faces corresponding to an $n$-slit experiment to be disjoint; if we know with certainty that the particle has passed through a particular slit, there should be no probability of finding it at another slit.

An $n$-slit experiment requires a system that has $n$ disjoint exposed faces $F_i, \ i\in\{1,...,n\}$. Consider an effect $(E|$ which represents the effect corresponding to the probability of finding a particle at a particular point on the screen. Then an $n$-slit experiment is a collection of effects $(e_I|,\ I\subseteq \{1,...,n\}$ such that \begin{equation} \label{n} (e_I|s)=(E|s),\hspace{3mm} \forall |s)\in F_I:=\bigcup_{i\in I} F_i,
\end{equation} and,
\begin{equation} \label{2} (e_I|s)= 0,\hspace{3mm} \forall |s) \text{ where }s\perp F_I.
\end{equation}
We can see these effects as being the composition of the transformation induced by closing the slits $\{1,...,n\}\setminus I$ and the effect $(E|$. If the particle was prepared in a state such that it would be unaffected by the slit closure (i.e. $|s)\in F_I$) then this composition should act the same as $(E|$ so that $(e_I|s)=(E|s)$. If instead the particle is prepared in a state which is guaranteed to be blocked (i.e. $|s')\perp F_I$) then we should obtain the zero effect so that $(e_I|s')=0$.

The relevant quantities for the existence of various orders of interference are therefore,
\begin{eqnarray}
& &I_1:=(E|s),\\
& &I_2:=(E|s)-(e_1|s)-(e_2|s),\\
& &I_3:=(E|s)-(e_{12}|s)-(e_{23}|s)-(e_{31}|s)+(e_1|s)+(e_2|s)+(e_3|s),\\
& &I_n:= \sum_{\emptyset\neq I\subseteq\{1,...,n\}}(-1)^{n-|I|}(e_I|s), \label{HOI}
\end{eqnarray}
for some state $|s)$ and defining $(e_{\{1,...,n\}}|:=(E|$. Where a theory has $n$th order interference if there exists a state $|s)$ such that $I_n\neq0$. Lack of third-order interference therefore means that the three slit interference pattern is the sum of the two-slit patterns minus the sum of the one-slit patterns. This is what we find for quantum theory. It was shown in \cite{sorkin1994quantum} that $I_n = 0 \implies I_{n+1}=0$, so if we have no $n$th order interference then there will be no $(n+1)$th order interference. It can be shown that classical probability theory satisfies $I_2=0$ and quantum theory satisfies $I_3=0$. The failure of $I_2=0$ for quantum theory means that the two-slit pattern is not just the sum of the one-slit patterns, this is just the usual notion of interference in the two-slit experiment.


\subsection{Requirements on a physical theory}

We have the following desiderata for a physically well-defined extension of quantum theory:
\begin{enumerate}

\item There should exist a well-defined state space\footnote{forming a convex cone in the GPT setting.} for an $N$-level system, $\Omega_N$ (for all finite $N$).

\item There should exist a well-defined effect space giving valid probabilities\footnote{in the GPT setting this will be a convex cone living inside the dual cone to the state space.}, $\mathcal{E}_N$.

\item Transformations should leave that state space invariant\footnote{In the GPT setting these will be linear maps that are completely preserving}.

\item Composite systems should be defined in a consistent way so that\footnote{Note that $\otimes$ here may not be the usual vector space tensor product \cite{barrett2007information, hardy2001quantum}.}, $\Omega_N\otimes\Omega_M=\Omega_{NM}$.

\item For a genuine extension of quantum theory there should exist a valid hyper-decoherence map.

\end{enumerate}


\section{Density cubes} \label{DC}

Daki{\'c} \emph{et al.} \cite{DensityCube} have proposed a method to construct a theory that exhibits third-order interference and extends standard quantum theory. They argue, based on the results in \cite{ududec2011three}, that the absence of third-order interference in quantum theory can be traced back to the fact that a quantum state coherently links at most two levels of the quantum system. This can be summarised as the fact that a quantum state is represented by a density matrix, where the matrix entries $\rho_{ij},\ i\neq j,$ are the coherences linking the levels $i$ and $j$. So in order for a theory to exhibit third-order interference the representation of states in said theory must contain terms that coherently link \emph{three} levels, i.e. terms of the form $\rho_{ijk}$, with $i,j,k$ all distinct. Thus a potential way to construct a theory that exhibits third-order interference is to consider a theory where the states are described not by matrices $\rho_{ij}$ as in quantum theory, but by (rank 3) \emph{tensors} with elements of the form $\rho_{ijk}$. Daki{\'c} \emph{et al.} refer to such tensors as \emph{density cubes}, as opposed to the density matrices of quantum theory.

\subsection{States and effects}

The basic features of the theory of density cubes are defined in analogy with quantum theory, as follows\footnote{See \cite{DensityCube} for a more comprehensive discussion.}. Every measurement outcome is associated with a density cube\footnote{i.e. the authors of \cite{DensityCube} require that their theory has a one-to-one correspondence between states and effects, in terms of GPTs this means that the state and effect cones are the same.} which, in general, has complex entries $\rho_{ijk}$. The element $\rho_{iii}$ is chosen to be real and corresponds to the probability of the outcome $i=1,...,n$ of a particular measurement. Thus $\sum_i\rho_{iii}=1$ and $\rho_{iii}\geq{0}$. In analogy to quantum theory, we refer to this property as the \emph{trace} of the density cube. In standard quantum theory the probability of finding the quantum state $\rho$ in the state $\sigma$ on measurement is given by  $p=Tr(\rho^{\dag}\sigma)=\rho_{ij}^*\sigma_{ij}$, where Einstein's summation convention has been adopted. In a similar manner, define $p=(\rho,\sigma)=\rho_{ijk}^*\sigma_{ijk},$ where $p$ denotes the probability of finding the a density cube in state $\rho$ when the measurement corresponding to the state $\sigma$ is applied. To ensure that $p$ is a real number, the constraint $\rho_{ijk}^*\sigma_{ijk}=\sigma_{ijk}^*\rho_{ijk}$ is enforced. In the quantum case $p\in\mathbb{R}$ is ensured as $\rho_{ij}$ is a Hermitian matrix, hence $\rho_{ij}=\rho_{ji}^*$. Similarly, call a density cube \emph{Hermitian} if exchanging two indices gives a complex conjugated element. As in the case of Hermitian matrices, Hermitian cubes form a real vector space with the inner product given by $(\rho,\sigma)=\rho_{ijk}^*\sigma_{ijk}$. We define pure states as those that satisfy the above conditions and also satisfy $(\rho,\rho)=1$. Positivity of the inner product, Hermiticity and the requirement that the terms $\rho_{iii}$ are probabilities are the only constraints imposed by \cite{DensityCube} on the structure of density cubes, and their state space.


For a three-level system, the normalization and Hermiticity conditions imply:
\begin{enumerate}\label{Herm}
\item $\rho_{iij}=\rho_{iij}^*=\rho_{iji}=\rho_{jii}, \quad i,j=1,2,3,\ i\neq j$,
\item $\rho_{123}=\rho_{312}=\rho_{231}=\rho_{213}^*=\rho_{321}^*=\rho_{132}^*$,
\item $\rho_{111}+\rho_{222}+\rho_{333}=1$,
\item $\rho_{iii}\geq{0}, \quad i=1,2,3$.
\end{enumerate}

Thus the density cube of a three-level system is specified by ten real parameters: point $1.$ contributes six real parameters (one for each choice of $i$ and $j$), point $2.$ contributes one complex, or two real, parameters, point $3.$ contributes three real parameters and point $4.$ reduces by one. This is two real parameters (one complex parameter) more than what is required to specify the state of a general three-level system (qutrit) in quantum theory. Thus, the elements $\rho_{ijk}$ with $i,j,k$ distinct can be seen as the crucial difference between the density matrix and the density cube. Therefore, based on the results in \cite{ududec2011three} discussed above, one might naively expect that the existence of the term $\rho_{ijk}$, with $i,j,k$ distinct, implies the existence of genuine third-order interference.


The complete characterisation of the density cube state space remains an important and interesting open problem. Nevertheless, some genuinely non-quantum density cubes were presented in \cite{DensityCube}. An example of such non-quantum density cubes (i.e. those with $\rho_{123}\neq 0$) are the following three pure states, first presented in \cite{DensityCube}:

$$\rho^{(j)}=\left\{ \begin{pmatrix} \frac{1-\delta_{1j}}{2}&0&0 \\ 0&0&\frac{\omega^{j-1}}{2\sqrt{3}} \\ 0&\frac{(\omega^{j-1})^*}{2\sqrt{3}}&0 \end{pmatrix}, \begin{pmatrix} 0&0&\frac{(\omega^{j-1})^*}{2\sqrt{3}} \\ 0&\frac{1-\delta_{2j}}{2}&0 \\ \frac{\omega^{j-1}}{2\sqrt{3}}&0&0 \end{pmatrix}, \begin{pmatrix} 0&\frac{\omega^{j-1}}{2\sqrt{3}}&0 \\ \frac{(\omega^{j-1})^*}{2\sqrt{3}}&0&0 \\ 0&0&\frac{1-\delta_{3j}}{2} \end{pmatrix}  \right\},$$

for $j=1,2,3$, where $\omega=e^{i\frac{2\pi}{3}}$ and $\delta_{ij}$ is the Kronecker delta. In each of the above density cubes, the element $\rho_{ijk}$ occurs in the $jk\mathrm{th}$ entry of the $i\mathrm{th}$ matrix in the list. It is easy to check that these density cubes are orthonormal, i.e. $(\rho_i,\rho_j)=\delta_{ij}$, and can be taken as part of a orthonormal basis in the real vector space of density cubes. We define a \emph{physical basis} as a set of density cubes that are orthogonal and sum to $\sum_n\delta_{in}\delta_{jn}\delta_{kn}$, these physical bases correspond to allowed (pure) measurements for density cubes.

\subsection{Transformations} \label{Transformations}

An example of a genuine `non-quantum' transformation between density cubes was also presented in \cite{DensityCube}. In order to present the constraints on transformations between density cubes imposed in \cite{DensityCube}, consider the following. Take the complex vector space of general rank-3 tensors, the Hermitian cubes, defined above, form a real subspace within this. A complex subspace can be defined by $\text{Span}[C^{(i)}]$ where $C^{(i)}$ are defined as,

$$C_{ijk}^{(n)}=\delta_{in}\delta_{jn}\delta_{kn}, \quad n=1,2,3,$$

$$C^{(k)}=\frac{1}{\sqrt{3}}\left\{\begin{pmatrix} 0&0&0 \\ 0&0&\delta_{4k} \\ 0&\delta_{5k}&0 \end{pmatrix}, \begin{pmatrix} 0&0&\delta_{5k} \\ 0&0&0 \\ \delta_{4k}&0&0 \end{pmatrix}, \begin{pmatrix} 0&\delta_{4k}&0 \\ \delta_{5k}&0&0 \\ 0&0&0  \end{pmatrix}  \right\}, \quad k=4,5,$$
note that $C^{(4)}$ and $C^{(5)}$ are not Hermitian cubes\footnote{A similar situation occurs in quantum theory: the Pauli matrices form a basis of the real vector space of Hermitian matrices, yet individual Pauli matrices are not physical states, only certain linear combinations of them are.} but the others are. A vector in $\text{Span}[C^{(i)}]$ is specified by five complex numbers. If we take the intersection of $\text{Span}[C^{(i)}]$ with the Hermitian cubes we obtain another real vector subspace where in the $C^{(i)}$ basis\footnote{We could instead use the basis which uses $C^{(4)}+C^{(5)}$ and $C^{(4)}-iC^{(5)}$ in which case our vectors would be written as five real numbers.} vectors are of the form $(p_1,p_2,p_3,z,z^*)^T,$ $p_i\in [0,1]\subset\mathbb{R}_+$, $z\in \mathbb{C}$ and with $\sum_{i=1}^3p_i=1$. This is a subspace of the Hermitian cubes. We must also impose our constraints as before, which gives the state space as a convex set living in this subspace.

The authors of \cite{DensityCube} consider only transformations that leave this subspace invariant. Aside from this the \emph{only} requirements imposed by the authors of \cite{DensityCube} are that the transformations are unitary matrices that map at least one physical basis of density cubes to another physical basis.

For example, consider a unitary transformation $T:D_0\rightarrow D$, where $D_0=\{q_1,q_2, q_3\}$ and $D=\{\rho_1, \rho_2, \rho_3\}$ are defined (in the $C^{(i)}$ basis) as follows,

$$\begin{aligned}
q_1&=(1,0,0,0,0)^T, \quad \rho_1=\frac{1}{2}(0,1,1,1,1)^T, \\
q_2&=(0,1,0,0,0)^T, \quad \rho_2=\frac{1}{2}(1,0,1,\omega, \omega^*)^T, \\
q_3&=(0,0,1,0,0)^T, \quad \rho_3=\frac{1}{2}(1,1,0,\omega^*,\omega)^T,
\end{aligned}$$
where as before $\omega=e^{\frac{2\pi i}{3}}$.

The $q_i$'s span a subspace of the `quantum states' of these density cubes. One matrix, provided by Daki{\'c} \emph{et al.}, that satisfies the conditions $Tq_i=\rho_i$, leaves this subspace invariant and is unitary is,

\begin{equation} \label{Transformation}
T=\frac{1}{2}\begin{pmatrix} 0&1&1&1&1 \\
                             1&0&1&\omega^*& \omega\\
                             1&1&0&\omega&\omega^*\\
                             1&\omega&\omega^*&1&0\\
                             1&\omega^*&\omega&0&1
                             \end{pmatrix}.
                             \end{equation}
Note that there are many matrices that satisfy the above condition, see \cite{DensityCube} for a more in-depth discussion.



\subsection{Hyper-decoherence} \label{hyper}


A hyper-decoherence mechanism will now be shown to exist in the theory of Density Cubes -- provided that there is an inner product preserving embedding {\color{black} (i.e. an injective, linear map)} of the quantum states into the density cube state space.


Such an embedding was given in \cite{DensityCube} and can be defined as follows. Denote an arbitrary quantum state by $\rho_{QT} {\color{black} \in \Omega_{QT}}$ and an arbitrary density cube by $\rho_{DC}{\color{black} \in\Omega_{DC}}$, where $\Omega_{QT}$ is the quantum state space, and so on. Define the embedding map $\mathcal{E}{\color{black} :\Omega_{QT}\to\Omega_{DC}}$ by:
$$ \big(\mathcal{E}[\rho_{QT}]\big)_{iii}=(\rho_{QT})_{ii}, \ \big(\mathcal{E}[\rho_{QT}]\big)_{iij}=\sqrt{\frac{2}{3}}\mathrm{Re}(\rho_{QT})_{ij}, \  \big(\mathcal{E}[\rho_{QT}]\big)_{ijj}=\sqrt{\frac{2}{3}}\mathrm{Im}(\rho_{QT})_{ij} \ \mathrm{for} \ i<j, $$
$$  \big(\mathcal{E}[\rho_{QT}]\big)_{ijj}=-\sqrt{\frac{2}{3}}\mathrm{Im}(\rho_{QT})_{ij} \ \mathrm{for} \ i>j, \ \mathrm{and} \  \big(\mathcal{E}[\rho_{QT}]\big)_{ijk}=0, \ \mathrm{for} \ i\neq j \neq k\neq i. $$
{\color{black} The other elements of the density cube are defined by the Hermiticity condition described in section~\ref{Herm}}
One can check that this embedding preserves the inner product. That is, we have that
$$(\rho_{QT},\sigma_{QT})_{QT}=(\mathcal{E}[\rho_{QT}], \mathcal{E}[\sigma_{QT}])_{DC},$$ where $(.,.)_{QT}$ is the inner product between quantum states, and so on.

{\color{black} To discuss hyper-decoherence it is useful to separate the density cubes into third order and lower order terms, we therefore write a generic density cube as, $$\rho_{DC}=\rho_{DC}^{(3)}+\rho_{DC}^{(2,1)}$$
where we define,
$$(\rho_{DC}^{(3)})_{ijk}:=\left\{\begin{array}{ll}(\rho_{DC})_{ijk} &\text{if } i\neq j\neq k\neq i \\ 0 &\text{otherwise}\end{array},\right.$$
 $$(\rho_{DC}^{(2,1)})_{ijk}:=\left\{\begin{array}{ll}0 &\text{if } i\neq j\neq k\neq i \\ (\rho_{DC})_{ijk} &\text{otherwise}\end{array}.\right.$$
 Note that $\rho_{DC}^{(2,1)}$ and $\rho_{DC}^{(3)}$ are not necessarily themselves valid density cubes.}

Given the above embedding, $\mathcal{E}$, one can define a hyper-decoherence map $\mathcal{D}$ as follows:
$$\mathcal{D}\circ\mathcal{E}=\mathds{1}_{QT}, \quad {\color{black} \mathcal{D}[\rho_{DC}^{(3)}]=0,}$$
  where $\mathcal{D}$ is a linear map\footnote{Note that linearity ensures that we can extend the map from the states on which it is defined to all Hermitian cubes.} from the real vector space of Hermitian cubes to the real vector space of Hermitian matrices, and $\mathds{1}_{QT}$ is the identity transformation on Hermitian matrices. This choice of $\mathcal{D}$ seems natural as we would expect such a map to leave any quantum state embedded in the Density Cube state space invariant and to eliminate the higher order coherences.

In order to show that $\mathcal{D}$ is a valid hyper-decoherence map we need to show that it maps all density Cube states to valid quantum states. That is $\mathcal{D}[\rho_{DC}]$ must be a positive, Hermitian operator with unit trace. That $\mathcal{D}[\rho_{DC}]$ has unit trace is guaranteed by the definition of $\mathcal{D}$ and the construction of the Density Cubes. To check the Hermiticity condition, consider the following. We have
\begin{equation} \label{hermiticity} \begin{aligned} (\mathcal{D}[\rho_{DC}])_{ij}&=\sqrt{\frac{{3}}{2}}\bigg(\big(\mathcal{E}\circ\mathcal{D}[\rho_{DC}]\big)_{iij} +i \big(\mathcal{E}\circ\mathcal{D}[\rho_{DC}]\big)_{ijj} \bigg), \ \mathrm{for} \ i<j \\
\mathrm{and}, \  (\mathcal{D}[\rho_{DC}])_{ij}&=\sqrt{\frac{{3}}{2}}\bigg(\big(\mathcal{E}\circ\mathcal{D}[\rho_{DC}]\big)_{iij} -i \big(\mathcal{E}\circ\mathcal{D}[\rho_{DC}]\big)_{ijj} \bigg), \ \mathrm{for} \ i>j.\end{aligned} \end{equation}
To show $\mathcal{D}[\rho_{DC}]^\dagger=\mathcal{D}[\rho_{DC}]$, we must check that $\big(\mathcal{D}[\rho_{DC}]\big)_{ij}=\big(\mathcal{D}[\rho_{DC}]\big)_{ji}^*$ for all $i,j$, but this follows from applying the Density Cube Hermiticity condition to equations~(\ref{hermiticity}).

To check the positivity property, we need to show that $\big(\mathcal{D}[\rho_{DC}], \sigma_{QT}\big)_{QT} \geq 0$, for all $\rho_{DC}$ and $\sigma_{QT}$.
Note that for suitable real coefficients $c_i$ we have that $\rho_{DC}^{(2,1)}=\sum_i c_i \mathcal{E}[\rho_{QT}^i]$, {\color{black} where $\rho_{QT}^i\in\Omega_{QT}$ are some arbitrary set of density matrices. We can }therefore write $\rho_{DC}=\sum_i c_i \mathcal{E}[\rho_{QT}^i]+\rho_{DC}^{(3)}$.
Combining this with the definition of $\mathcal{D}$, we have that
$$\begin{aligned}
\big( \mathcal{D}[\rho_{DC}], \sigma_{QT}\big)_{QT} =& \big( \mathcal{E}\circ\mathcal{D}[\rho_{DC}], \mathcal{E}[\sigma_{QT}]\big)_{DC} \\
=& \big( \mathcal{E}\circ\mathcal{D}\big[\sum_i c_i \mathcal{E}[\rho_{QT}^i]+\rho_{DC}^{(3)}\big], \mathcal{E}[\sigma_{QT}]\big)_{DC}  \\
=& \big( \sum_i c_i \mathcal{E}[\rho_{QT}^i], \mathcal{E}[\sigma_{QT}]\big)_{DC} \\
=& \big( \rho_{DC}, \mathcal{E}[\sigma_{QT}]\big)_{DC} - {\big(\rho_{DC}^{(3)}, \mathcal{E}[\sigma_{QT}]\big)_{DC}} \\
=& \big( \rho_{DC}, \mathcal{E}[\sigma_{QT}]\big)_{DC} \\
 \geq & 0.
\end{aligned}$$
The equation $\big( \mathcal{D}[\rho_{DC}], \sigma_{QT}\big)_{QT} =\big( \rho_{DC}, \mathcal{E}[\sigma_{QT}]\big)_{DC}$, derived above, implies that the embedding map $\mathcal{E}$ is the adjoint of the hyper-decoherence map $\mathcal{D}$. This may prove useful in further constructions of higher-order interference theories.

Given the embedding $\mathcal{E}$, the hyper-decoherence map defined above maps density cubes to valid quantum states. One should note however, that the existence of this embedding is not guaranteed by the axioms of the Density Cube framework, but is a very reasonable constraint if one wants an extension of quantum theory.

%
In quantum theory, to have coherence between two levels of the quantum state described by the density matrix $\rho_{ij}$ there must be some probability of finding the state in either of the levels that the coherence is between. These probabilities set a bound on the degree of coherence possible, e.g. for a qubit we have $|\rho_{01}|^2\leq\rho_{00}\rho_{11}$. Based on this, one might expect that any third order coherence in a Density Cube would be supported by second and first order coherences. Interestingly this is not the case in the Density Cube framework; states in the physical basis $D$ considered by Dak{\'i}c \emph{et al.} have third order terms but all second order terms are zero. While it is the case that the positivity condition imposes some bounds on the higher-order coherences, there may be further constraints that need to be imposed to have a well-defined theory. It would be interesting if future constructions of higher-order interference theories had this property.



\subsection{A computational advantage?} \label{Advantage}

When comparing quantum theory with other foil theories, an approach that has proved fruitful in recent years is to compare their performance in information-theoretic tasks. We will now show that the theory of Density Cubes has a slight advantage over quantum theory in a computational task we call the `three collision problem', which is a variation of the standard collision problem discussed in \cite{collision}. The three collision problem is defined as follows: given a function from a trit to a bit, $f:\{0,1,2\}\rightarrow\{0,1\}$, determine if $f(0)=f(1)=f(2)$. As is standard in quantum computation, we represent this problem with a black-box oracle.  Performance will be measured via the probability of error after a single query to this oracle, given the caveat that if $f(0)=f(1)=f(2)$ there must be zero error.

Let $\{\ket{i}\}$, for $i=0,1,2$, be the quantum computational basis and consider the following quantum oracle for this problem:
$$\mathcal{O}_f^{QT}\ket{i}=(-1)^{f(i)}\ket{i}.$$
This oracle is the same as the one considered by Grover in his search algorithm \cite{NC}, and it is easy to check it is unitary. Preparing a superposition over the three basis states and querying the oracle leaves us in the state
$$\frac{1}{\sqrt{3}}\big((-1)^{f(0)}\ket{0}+(-1)^{f(1)}\ket{1}+(-1)^{f(2)}\ket{2}\big).$$
If $f(0)=f(1)=f(2)$, then the state, up to a global phase, is: $\frac{1}{\sqrt{3}}\big(\ket{0}+\ket{1}+\ket{2}\big),$ while if they are not equal the state, up to a global phase, is one of: $\frac{1}{\sqrt{3}}\big(-\ket{0}+\ket{1}+\ket{2}\big), \frac{1}{\sqrt{3}}\big(\ket{0}-\ket{1}+\ket{2}\big),$ or $\frac{1}{\sqrt{3}}\big(\ket{0}+\ket{1}-\ket{2}\big).$ As the state with $f(0)=f(1)=f(2)$ is not orthogonal to the other three, there does not exist a measurement that can perfectly distinguish them and the error after one query is $1/9$.

Dak{\'i}c \emph{et al.} have provided a way to associate one of three density cubes to a pure three-level quantum state $\ket{\psi}=c_0\ket{0}+c_1\ket{1}+c_2\ket{2}$. The association is as follows:
$$
\begin{aligned}
\rho_{iij}^{(n)}=-\frac{1}{\sqrt{6}}\mathrm{Re}(c_i^*c_j), \quad &\rho_{ijj}^{(n)}=-\frac{1}{\sqrt{6}}\mathrm{Im}(c_i^*c_j) \ \mathrm{for} \ i<j, \quad \rho_{iii}^{(n)}=\frac{1}{2}(1-|c_i|^2), \\ &\mathrm{and}, \quad \rho_{012}^{(n)}=\frac{\omega^n}{2\sqrt{3}} \ \mathrm{with} \ n=0,1,2.
\end{aligned}
$$
{\color{black} The other elements of the density cube are determined by the Hermiticity condition (see section \ref{Herm}).} Note that only the third-order terms depend on the value of $n$. One can show \cite{DensityCube} that $(\rho^{(n)}(\ket{\phi}), \rho^{(m)}(\ket{\psi}))_{DC}=\frac{1}{4}(1+|\langle\phi | \psi\rangle|^2 )+\frac{1}{2}\cos\frac{2\pi(n-m)}{3}\geq 0$. Given this association, we can describe a Density Cube oracle for the three collision problem as follows. The oracle acts as the
{\color{black} quantum oracle on the `quantum part' of the density cube, but also acts on the `higher order term' i.e. the value of $n$. We define the oracle as, $$\mathcal{O}^{DC}_f::\rho^{(n)}(\ket{\psi})\mapsto \rho^{(n_f)}(\mathcal{O}^{QT}_f\ket{\psi})$$ where $n_f=n+f(0)+f(1)+f(2)=n+\sum_if(i)$.} One can check that the action of this oracle leaves the fragment given by the above association invariant. While it may appear odd at first to allow density cubes with non-zero higher-order terms access to the value $f(0)+ f(1)+f(2)$, it should be noted that quantum theory has a similar advantage over classical theory when accessing a computational oracle. In classical computing one can only access the value of $f$ on a single value $i$ per query of an oracle, but, in quantum theory, one can access information about $f(i)+f(j)$ by querying the same oracle in superposition. It thus seems reasonable to allow third-order terms $\rho_{012}^{(n)}$ access to information about the value of $f(0)+ f(1)+f(2)$.

Let $\ket{\phi}=\frac{1}{\sqrt{3}}\big(\ket{0}+\ket{1}+\ket{2}\big),$ and prepare the density cube $\rho^{(0)}(\ket{\phi})$. Applying the Density Cube oracle leaves this state invariant if $f(0)=f(1)=f(2)$ and maps this state to either $\rho^{(1)}(\mathcal{O}^{QT}_f\ket{\phi})$ or $\rho^{(2)}(\mathcal{O}^{QT}_f\ket{\phi})$ otherwise, thus giving an error probability of $$(\rho^{(0)}(\ket{\phi}), \rho^{(1)}(\mathcal{O}^{QT}_f\ket{\phi}))=(\rho^{(0)}(\ket{\phi}), \rho^{(2)}(\mathcal{O}^{QT}_f\ket{\phi}))= 1/32$$ after a single query. The theory of Density Cubes thus provides a slight advantage over quantum theory in the three collision problem.

\subsection{Issues with the Density Cube framework} \label{DC1}
In this section two possible issues with the framework of Density Cubes will be presented and discussed. In particular it will be demonstrated that  the axioms imposed in defining the theory are insufficient to uniquely characterise the state space. We also show that the definition of transformations employed by \cite{DensityCube} allows for transformations in the theory that map well-defined states to density cubes that give complex-valued probabilities for certain measurement outcomes.

\subsubsection{Axioms insufficient to specify a unique operational theory}

Daki\'{c} \emph{et al.} mention that they have not fully constructed the state space for density cubes \cite{DensityCube}, instead they present a particular set of states which satisfy their axioms (i.e. they are Hermitian, have unit trace and are each positive with respect to the others). The difficulty in fully constructing the state space stems from the positivity axiom. In quantum theory we can define positivity as, \[Tr(\rho^\dagger \sigma)\geq 0\quad \forall \rho,\sigma,\] where $\rho$ and $\sigma$ are density matrices. This is analogous to the positivity condition imposed by Daki\'{c} \emph{et al.} and we refer to this property as `relative positivity'. In practice this is a difficult property to use to construct a state space, there is -- potentially -- an infinite number of conditions to check for each state in the state space. In quantum theory we can avoid this problem by using an alternative -- and equivalent -- definition of positivity, that \[\forall \lambda \in \mathsf{Eigenvalues}(\rho)\quad \lambda \geq 0.\] This is a single state property rather than a relative property and so it is simple to construct a state space by imposing this condition. There is no equivalent condition for density cubes as we do not have any eigenvalues for rank 3 tensors, we are therefore limited to using relative positivity.

Given that we only have a relative notion of positivity, it is possible to construct different state spaces depending on which set of states we choose to start with. However we know that -- if we want a genuine extension of quantum theory -- we need some (Hermitian, trace and inner product preserving) embedding of the quantum states into the Density Cube state space. Daki\'{c} \emph{et al.} present one such embedding, which we will discuss in more detail in section \ref{hyper}. It is conceivable that given such an embedding the state space is uniquely specified, and that this choice of embedding is analogous to a choice of re-parametrization of the quantum state space, and, as such, leads to operationally equivalent theories. Unfortunately this is not the case; it is possible to construct operationally distinct theories within the Density Cube framework. As such, the axioms imposed are not sufficient to uniquely characterise the theory.

For example, consider the embedding of quantum states described in \cite{DensityCube}, discussed in section \ref{hyper}, and use the basis $\{C^{(i)}\}$ described above. Then we can consider the states \[c=\frac{1}{2}(1,1,0,1,1)^T\ \text{and}\ v=\frac{1}{256}\left(10,10,236,-\left(65+i\sqrt{595}\right), -\left(65-i\sqrt{595}\right)\right)^T,\] these are both quantum states with added higher-order coherence terms and so will be positive with respect to all of the quantum states. However they are not positive with respect to each other, $(c,v)<0$. There is no reason to prefer one of these to the other and we cannot add both to the state space, we therefore have an arbitrary choice at this stage in how to construct the theory. Note that both of these states are positive with respect to the physical basis $D$, but have different inner products with elements of $D$. Thus choosing which state to include in the state space will lead to theories that make operationally distinct predictions about certain measurements.

Given the above discussion it may therefore be better to consider the theory of density cubes more as a framework for developing theories in. The (partial) state space of Daki\'{c} \emph{et al.} would then be one example of a state space within this framework. The difficulty in constructing the complete state space causes further problems when defining transformations within the theory. In most GPTs -- given that there is a complete geometric view of the state space -- it is simple to define transformations as linear transformations that map the state space into itself. However, if we are not given a complete state space it is not possible to define transformations in this way.

\subsubsection{{\color{black} Characterising the set of physical transformations} \label{DCNonPhys}}

{\color{black}
We will now show that the lack of fully constructed state space is also problematic for defining allowed transformations within the theory. Daki{\'c} \emph{et al.} present a particular transformation $T$ that they use throughout their paper. It can be shown that for the particular fragment that they are constructing that this is a valid transformation. By valid transformation we mean that it is linear and maps states to states. They also provide a set of axioms which need to be satisfied such that a transformation is valid. We show that these are necessary but not sufficient conditions as we will demonstrate.

The axioms that they impose in \cite{DensityCube} -- as we discussed in section~\ref{Transformations} -- are,
\begin{enumerate}
\item linearity
\item unitarity
\item subspace preserving
\item map between physical bases (e.g. $D_0\mapsto D$).
\end{enumerate}
This allows for transformations such as $T'$, eq.~\ref{TNew}, which can easily be shown to violate the Hermiticity of states.
 \begin{equation} \label{TNew}
T'=\frac{1}{2}\begin{pmatrix} 0&1&1&\frac{1}{2}(1+\sqrt{3})&\frac{1}{2}(-1+\sqrt{3}) \\
                             1&0&1&\frac{1}{2}(\omega^*+\sqrt{3}\omega)& \frac{1}{2}(-\omega+\sqrt{3}\omega^*)\\
                             1&1&0&\frac{1}{2}(\omega+\sqrt{3}\omega^*)& \frac{1}{2}(-\omega^*+\sqrt{3}\omega)\\
                             1&\omega&\omega^*&\frac{1}{2}&\frac{\sqrt{3}}{2}\\
                             1&\omega^*&\omega&\frac{\sqrt{3}}{2}&-\frac{1}{2}
                             \end{pmatrix}.
                             \end{equation}
                             For example, $T'(\rho_1)$ has complex elements that should be real and so provide complex probabilities.

                            $$\begin{aligned} T'(\rho_1)=&\Bigg(\frac{1}{2}+\frac{\sqrt{3}}{4}, \frac{1}{4}\left(1-\sqrt{3}\left(\frac{1+i}{2}\right)\right) ,\frac{1}{4}\left(1-\sqrt{3}\left(\frac{1-i}{2} \right)\right),\frac{\sqrt{3}-1}{8} ,\frac{\sqrt{3}-3}{8} \Bigg)^T \\
\approx&\large(0.9,0.03 -0.2 i,0.03+0.2 i,0.09,-0.2\large)^T. \end{aligned} $$
 The usual solution to this would be to require that transformations map states to states or equivalently that they preserve Hermiticity and positivity, which would rule out `unphysical' transformations such as $T'$.

Using Hermiticity and positivity preservation as a characterisation of transformations however is dependent on the state space, and, as we do not have a complete state space, these are impossible to enforce in practice. Characterising transformations beyond the specific example of Daki{\'c} \emph{et al.} is not possible at this stage.


This again highlights the issue that different fragments give operationally distinct predictions, we see here that not only the possible states depend on the choice of state space but that the set of physical transformations depends on this choice as well.

}

\section{Quartic quantum theory}\label{QQT}

Quartic quantum theory (QQT) was developed by {\.Z}yczkowski \cite{zyczkowski2008quartic} as an attempt to realise the $K=N^4$ level of the tomographic hierarchy introduced by Hardy in \cite{hardy2001quantum}. This is to be contrasted to quantum theory which satisfies\footnote{This is allowing for subnormalised states hence quantum theory having $K=N^2$ rather than $K=N^2-1$.} $K=N^2$ and classical probability theory which satisfies $K=N$. Density cubes however satisfy $K=N^2+2{N\choose 3}\neq N^r$ and so are not in Hardy's hierarchy\footnote{Thus implying that the theory of Density Cubes violate Hardy's principle of tomographic locality, which roughly says that composite states can be characterised by local measurements.}.

We have discussed the connection between tomography and higher-order interference presented in \cite{ududec2011three}, specifically, the dimension of the subspace on which one must perform measurements to do complete tomography corresponds to the order of interference.
In the $K=N^r$ hierarchy, post-quantum theories require tomography on greater than two dimensional subspaces. In light of this, QQT provides a potential candidate for a GPT that exhibits higher order interference.

\subsection{Description of the theory}
We will provide a brief overview of the theory here but refer to the original paper \cite{zyczkowski2008quartic} for the details. The state space for an $N$ level QQT system is constructed from a restriction of an $N^2$ level quantum system (i.e. the tensor product of two $N$ level quantum systems). The restriction limits us to convex combinations of states that are unitarily connected to\footnote{Where we are using curved brackets to denote QQT states and effects and Dirac brakets to represent the underlying quantum density matrix description of the state} $|s_{initial}):=\frac{1}{N}\mathbb{I}\otimes \left|0\right>\left<0\right|$, so the state space is given by the convex hull of $|s)\in\{U(\frac{1}{N}\mathbb{I}\otimes \left|0\right>\left<0\right|)U^\dagger \hspace{2mm}|\hspace{2MM}U\in SU(N^2)\}$, i.e. we allow for arbitrary mixtures of any states which can be reached by applying arbitrary unitaries to the composite quantum system beginning in state $|s_{initial})$.

The restriction on the quantum state space essentially imposes that there is a maximum purity that the state can reach. This can be seen to be roughly analogous to the epistemic restriction used by Kochen and Specker in their hidden variable model \cite{Kochen-Specker} and also by Spekkens in his Toy Model \cite{spekkens2007evidence}, where the state space of a two level system is given by a pair of classical bits (i.e. a pair of two level classical systems) with a restriction imposed on how much one can know about the system\footnote{It may be illuminating to consider what effect imposing an epistemic restriction has on the structure of arbitrary GPTs and then to view Spekkens Toy Model and Quartic Quantum Theory as special cases of this.}.

Transformations are defined as being linear maps that leave the state space invariant and which are completely preserving, i.e. $T:\Omega_N\to\Omega_N$ and $T\otimes \mathbb{I}_M:\Omega_{NM}\to\Omega_{NM} \hspace{3mm} \forall M$, where $\Omega_N$ is the QQT state space for an $N$ level system. The last condition is a generalisation of complete positivity in quantum theory.

Effects satisfy the `no-restriction hypothesis~\cite{LalJanotta}' which says that any mathematically well-defined effect is allowed. That is an effect $(e|$ is allowed in the theory if it is linear and $0\leq (e|s) \leq 1, \forall |s) \in \Omega_N$. We have imposed a restriction on the Quantum theory state space\footnote{Geometrically the unnormalised state space of any GPT is a convex cone (the normalised state space is the intersection of a hyperplane with the convex cone), and if the no-restriction hypothesis is satisfied then the effect space is the dual cone. If the state space is restricted this increases the size of the dual cone, this is what we find in QQT.}, and so the effect space is enlarged. For example we can have effects such as $(e|=N\left|0\right>\left<0\right|\otimes\left|0\right>\left<0\right|$. Which in quantum theory could give probabilities greater than one, but due to the restriction on purity this cannot happen in QQT. This is because $(e|s)\leq N\lambda_s^{max}$. Where $\lambda_s^{max}$ is the maximum eigenvalue of the density matrix representation of the state $|s)$. For a QQT state $\lambda_s^{max}\leq 1/N$ so $(e|S)\leq N/N = 1$.

This fully constructs the theory for an $N$ level quartic quantum system as we have a complete consistent description of all of the states, transformations and effects that exist in the theory. There is also a consistent notion of hyper-decoherence by which any quartic quantum system can decohere to a quantum system. In QQT decoherence is represented by a partial trace over one of the quantum sub-systems, this clearly can only map us to the quantum state space, additionally any quantum state, $\rho$, can be reached in this way through $\rho = Tr_{A2}(\rho\otimes\frac{1}{N} \mathbb{I}_{A2})$. {\color{black} It is worth reiterating that these are not physical subsystems. The choice of tensor product decomposition and which part to trace out is therefore entirely arbitrary.}

\subsection{Interference in Quartic Quantum Theory} \label{interference}

We will consider interference in the context of definitions 19 and 20 in \cite{barnum2014higher}, which were described in section~\ref{GPT}, and show that QQT has $N$th order interference in an $N$ level system. Firstly we define the faces,$$F_i:=\left\{\frac{1}{N}\left|i\right>\left<i\right|\otimes\mathbb{I}\right\}=\left\{\frac{1}{N}\sum_{j=1}^N\left|ij\right>\left<ij\right|\right\},$$ we can then choose a set of effects which satisfy the constraints given by the definition. These are,
\begin{eqnarray}\label{QQTEffects}
(E| &:=& \sum_{i,j=1}^N\left|ij\right>\left<ij\right|, \\
(e_i| &:=& N\left|ii\right>\left<ii\right|, \nonumber \\
(e_I| &:=& \sum_{i\in I}\sum_{j=1}^N \left|ij\right>\left<ij\right|, \text{ for } I\subseteq\{1,...,N\}.\nonumber
\end{eqnarray}
It is simple to show that these do satisfy Eq.~(\ref{n}) and Eq.~(\ref{2}), for instance if $|s)\in F_i$, we have $(e_i|s)=\mathrm{Tr}((N\left|ii\right>\left<ii\right|) (\frac{1}{N}\left|i\right>\left<i\right|\otimes\mathbb{I}))=1=(E|s)$ and if $|s')\perp F_i$, we have $(e_i|s')=\mathrm{Tr}((N\left|ii\right>\left<ii\right|)(\frac{1}{N}\left|j\right>\left<j\right|\otimes\mathbb{I}))=0$, as required.

If we consider a three level system, $N=3$, then we have third-order interference if,
$$ (E| \neq \sum_{i>j}(e_{\{i,j\}}| - \sum_i (e_i|, $$
which is the case here. We have,
$$ (E| = \sum_{i,j=1}^N\left|ij\right>\left<ij\right| \neq 2 \sum_{i\neq j }\left|ij\right>\left<ij\right| - \sum_i\left|ii\right>\left<ii\right|=\sum_{i>j}(e_{\{i,j\}}| - \sum_i (e_i|$$

If instead of $(e_i|$ we used the effects $(e_I|$ where $I=\{i\}$, these also satisfy the conditions in the definition but don't give us third-order interference, that is
$$(E| = \sum_{i,j=1}^N\left|ij\right>\left<ij\right| = \sum_{i>j}(e_{\{i,j\}}| - \sum_i (e_{\{i\}}|.$$ So we see that we obtain higher order interference by using the super-quantum effects allowed in QQT.

It can be shown that this approach generalises to Nth order interference. That is that we can find a set of effects such that,
$$(E| \neq \sum_{\emptyset \neq I\subseteq \{1,...,N\}} (-1)^{N-|I|} (e_I|.$$ A valid set of effects for this are those defined above in Eq.(\ref{QQTEffects}), where we see Nth order interference if we replace $(e_{\{i\}}|\to(e_i|$. The simplest way to see this is to observe that the effects $(e_I|$ are all the quantum effects for a $N$-slit experiment tensored with the identity, therefore using these we will not see higher-order interference, but if we replace the `quantum' $(e_{\{i\}}|$ with a super-quantum $(e_i|$ and note that $\sum_i(e_i|\neq \sum_i(e_{\{i\}}|$ then we will see higher order interference to all orders.

Note that we obtain this result as the constraints imposed in the definition of higher-order interference are insufficient to uniquely determine the effects $(e_I|$, this is perhaps a problem with the definition. If one were to actually perform the experiment then there should be a unique description of the effect corresponding to what happens, as, at the operational level, it should arise from blocking slits in a physical barrier. Thus for the definition of Barnum \emph{et al.} to correspond to this physical picture in an operationally meaningful way, extra constraints must be imposed on the theory under consideration.

Based on the above discussion, one should not consider QQT as an example of a theory that exhibits higher-order interference in the sense originally meant by Sorkin, but rather a demonstration of the challenges of applying his original definition to arbitrary GPTs. Thus to begin to understand the reason why, in some sense, quantum theory is limited to second-order interference, we first need a definition of higher-order interference that is applicable to, and makes good operational sense in, arbitrary GPTs. Ways in which such a definition might arise will be discussed in section \ref{con}.

\subsection{Composite systems in Quartic Quantum Theory} \label{QQTComp}

The main limitation of quartic quantum theory {\color{black} -- as discussed by {\.Z}yczkowski \cite{zyczkowski2008quartic} --} is that it does not deal with composite systems. The difficulty with defining composite systems is ensuring that discarding part of a composite system does not result in a state outside the (single system) QQT state space. For example if we define composition in the same way as quantum theory then a bipartite QQT system would be made of four quantum systems, two of which are required to be in the maximally mixed state. If we then allow for arbitrary quantum transformations on this system then we can use a swap unitary to put all of the mixed systems into one half of the bipartition and all of the pure systems into the other. If we then discard the mixed partition we are left with a pure quantum state, which is not a valid state in QQT. In other words, marginalisation takes us outside the QQT state space.

For example, if we prepare the state $ |s_{AB}) = \frac{1}{N^2}\left|0\right>\left<0\right|_{A1}\otimes\mathbb{I}_{A2}\otimes\left|0\right>\left<0\right|_{B1}\otimes\mathbb{I}_{B2},$
apply a swap to the middle two systems ($A2$ and $B1$),
$ U_{swap}^{A2,B1}|s_{AB}) =\frac{1}{N^2} \left|0\right>\left<0\right|_{A1}\otimes\left|0\right>\left<0\right|_{A2}\otimes \mathbb{I}_{B1}\otimes\mathbb{I}_{B2},$
then discarding system $B$ gives,
$\left|0\right>\left<0\right|_{A1}\otimes\left|0\right>\left<0\right|_{A2},$ which is outside the state space as it is `too pure'.

A possible solution to this problem is to impose a restriction on the allowed transformations to try to avoid a situation like this. For example, allowing only separable transformations would mean that it was impossible to apply the swap between the two QQT systems and so discarding one of them could not cause problems. This would mean that there were no entangling dynamics in the theory\footnote{Note that another commonly studied GPT, known as `box world', also shares this feature \cite{short2010strong}.} and that we are unable to reversibly prepare an entangled state from a product state, amongst other things. An interesting direction to pursue would be whether this can be seen as a consequence of third-order interference, or whether it is possible to have a theory with third-order interference and similar entangling dynamics to those that we have in quantum theory.

%
%

\subsubsection{Note on Boxworld-like correlations in Quartic Quantum Theory} \label{Correlation}
In using the quantum tensor product in the previous section we are relying on a commonly used axiom in quantum reconstructions, that any $N$ level system should be equivalent, i.e. a single system with $N$-levels should be equivalent to a composite system that has $N$-levels. If we relax this assumption then we can instead use some other tensor product\footnote{This tensor product will have to give a state space bound by the minimal and maximal tensor products, see \cite{LalJanotta} for details.}.

We note that if we consider the `classical' subspace of a two-level quartic quantum system, i.e. the diagonal density matrices, the state space corresponding to these states forms an octahedron \cite{zyczkowski2008quartic}, and the effect space dual to this forms a cube. This is the `unrestricted Spekkens Toy Model' state and effect space discussed in \cite{LalJanotta}. Janotta and Lal discuss how the (generalised) maximal tensor product of such a state space gives rise to PR box correlations, i.e. those that maximally violate a Bell inequality whilst maintaining no-signalling. We therefore should be able to obtain the same correlations if we take the maximal tensor product of two two-level quartic quantum systems. Such correlations imply a speed-up over quantum theory in communication complexity problems and this opens the door to investigations of the information processing power of well-defined physical theories with higher-order interference.

We have seen that QQT is a well-defined extension of quantum theory and so may prove a useful foil in understanding the certain features of the quantum formalism.

\section{Conclusion} \label{con}

This paper considered two proposed extensions of quantum theory: Daki{\'c} \emph{et al.}'s Density Cubes and {\.Z}yczkowski's Quartic Quantum Theory. Our investigation clarifies the impact of these two generalised theories to ongoing experimental tests for higher-order interference and explores potential information-theoretic consequences of post-quantum interference in these concrete theories. We examined their order of interference relative to the hierarchy defined by Sorkin and investigate whether these theories satisfy natural physical conditions one would expect from an extension of quantum theory. Our results are summarised in the table below.

\begin{center}
\vspace{.5mm}
\begin{minipage}[c]{\paperwidth}
\begin{tabular}{|l|c|c|}
  \hline
 \textbf{Desiderata}    & \textbf{Density cubes} & \textbf{Quartic quantum theory} \\ \hline

  States $s\in\Omega_N$ & $\checkmark$\footnotemark[1]  & $\checkmark$ \\ \hline

  Effects $e \in \mathcal{E}_N \subseteq \Omega_N^*$ & $\checkmark$ & $\checkmark$ \\ \hline

  Transformations $T:\Omega_N\rightarrow\Omega_N$ & ?\footnotemark[2] & $\checkmark$ \\ \hline

  Composite systems $\Omega_N\otimes\Omega_M=\Omega_{NM}$  & $\times$ & ?\footnotemark[3] \\ \hline

  Higher order interference i.e. $n>2$ in Eq.~(\ref{HOI}) & $\checkmark$  & ?\footnotemark[4] \\
  \hline
  Hyperdecoherence & $\checkmark$ & $\checkmark$\\ \hline
\end{tabular} \\
{\small  $^{1}$ But not uniquely fixed by the constraints in the theory, see Sec. \ref{DC}.\\
$^2$ We show in Sec. \ref{DCNonPhys} that transformations -- as defined in \cite{DensityCube} -- take us out of the state space.\\
$^3$ We suggest a definition of composite systems by limiting to only local transformations, see Sec. \ref{QQTComp}.\\
$^4$ $n$th order interference, for all $n$. This is a result of a deficiency in the definition of higher-order int-\\ erference, see Sec. \ref{QQT}.\\ }
\end{minipage}
\end{center}

The specific partial state space and single transformation presented in \cite{DensityCube} do indeed exhibit third-order interference. However this state space is not uniquely specified by the imposed axioms, and there exist other transformations allowed by these axioms which lead to unphysical results. We therefore suggest that it would be interesting to investigate what further axioms would be necessary to uniquely specify a state space, as such a construction would provide a natural way of characterising the physical transformations. We showed that, if one has an embedding of quantum theory into a specific Density Cube state space, the adjoint of this embedding gives a suitable hyper-decoherence mechanism. Considering further consistency requirements with quantum theory may help with fully developing the theory and may provide a complete axiomatisation. Given this, one could compare the Density Cube theory to quantum theory in a rigorous manner and hope to learn in what ways the theories differ, thus taking a step toward a better understanding of what it means to live in a quantum world.

The operational definition of higher-order interference of Barnum \emph{et al.} suffers from an ambiguity; the specification of the effect $(E|$ does not uniquely fix the effects $(e_I|$ in an arbitrary GPT\footnote{If the GPT in question supports filters \cite{barnum2014higher}, then the effects can be uniquely specified by a choice of filters. As this is the only situation of interest to Barnum \emph{et al.}, their definition suffices for all considerations of interest in \cite{barnum2014higher}.} as can be seen in Sec. \ref{QQT}. We would intuitively expect that once $(E|$ is specified the effects $(e_I|$ are fixed, as they should arise from blocking a certain number of slits in a physical barrier. Thus to begin to illuminate \emph{why}, in some sense yet to be defined,\footnote{For example, the computational power of different GPTs has recently been investigated \cite{LeeBarrett, LeeHoban, LeeSelby} and it could be the case that higher-order interference implies implausible computational power, or indeed, trivial computational power. If either of these were the case, it could be seen as the `reason' why quantum theory does not exhibit higher-order interference.} quantum theory is limited to only second-order interference, we first need a definition of higher-order interference that is applicable to, and makes good operational sense in, arbitrary GPTs.
 
In quantum theory there is an intimate relation between interference and phase, which is illustrated most clearly by the Mach-Zender interferometer. The connection between phase and interference is not touched on by the Barnum \emph{et al.} notion of higher-order interference. Garner \emph{et al.} \cite{Garner} have proposed a definition of phase and interference applicable to an arbitrary GPT, but their definition of interference bears no resemblance to Sorkin's hierarchy and as such they do not discuss higher-order interference. The subject of phase transformations and higher-order interference is being investigated \cite{LSprep} and may result in a definition of higher-order interference that is applicable to arbitrary GPTs.

It was shown in \cite{Interference-speed-up} that quantum interference is necessary for a quantum computer to be hard to classically simulate. It is thus interesting to note that there are indications in the theories discussed here that higher-order interference gives an advantage over quantum theory in certain information-theoretic tasks. However, it remains to be seen whether this can be shown to be a direct consequence of the existence of higher-order interference or whether it is due to other features of the theories.




\subsection*{Acknowledgements}
The authors thank M. Hoban and J. Richens for valuable feedback on an earlier draft of the current paper. This work was supported by the EPSRC through the Quantum Controlled Dynamics Centre for Doctoral Training and the University of Oxford Department of Computer Science. CL also acknowledges funding from University College, Oxford.

\end{document}